\documentclass[aps,prd,twocolumn,superscriptaddress,longbibliography]{revtex4-2}
\usepackage{amsmath,amssymb}
\usepackage{graphicx}
\usepackage{hyperref}
\usepackage{booktabs}
\usepackage{xcolor}
\usepackage{siunitx}

\newcommand{\NRSur}{\textsc{NRSur7dq4}}

\begin{document}

\title{A fast, differentiable neural-network surrogate\\
       for precessing binary black-hole waveforms}

\author{Beka Modrekiladze}
\affiliation{Deutsches Elektronen-Synchrotron DESY, Notkestra{\ss}e 85, 22607 Hamburg, Germany}
\affiliation{Carnegie Mellon University, Pittsburgh, PA 15213, USA}

\begin{abstract}
Gravitational-wave parameter estimation requires millions of waveform
evaluations per event, a cost that constrains real-time inference and
population studies.
We present a fast, fully differentiable neural-network surrogate for the
precessing numerical-relativity model \NRSur{}, spanning its full
intrinsic parameter space
$\lambda=(q,\,\boldsymbol{\chi}_1,\,\boldsymbol{\chi}_2)$
together with the reference orbital frequency $\omega_0$.
Rather than a single polarization at a fixed orientation, the surrogate
predicts the \emph{inertial-frame spherical-harmonic modes} $h_{\ell m}$
($\ell\leq 4$), so that both polarizations $h_+,h_\times$ at an arbitrary
orientation $(\iota,\varphi,\psi)$ are reconstructed from one network
evaluation through an analytic, differentiable mode-to-strain projection.
Trained on $6\times10^5$ waveforms, it attains a fixed-orientation match
of mean $0.975$ (median $0.988$) and an orientation-averaged match of
mean $0.940$ (median $0.975$) for $q\in[1,4]$, $|\boldsymbol{\chi}_{1,2}|\leq 0.8$,
while keeping the overall strain amplitude physical (median ratio $0.98$).
It generates a waveform in $\SI{12}{ms}$ (single) and ${\sim}3.5\times10^4$
per second in batches on a single GPU. Because the mode-to-strain projection is analytic, the surrogate is
differentiable in both intrinsic and extrinsic parameters, yielding a full
13-dimensional Fisher matrix (validated against finite differences) and
gradient-based (HMC/NUTS) parameter estimation.
\end{abstract}

\maketitle

\section{Introduction}
\label{sec:intro}

The detection of gravitational waves (GWs) from merging compact
binaries~\cite{LVK:GW150914} by the Advanced LIGO and Virgo
detectors~\cite{aLIGO:2015,AdVirgo:2015} rests on matched filtering:
comparing detector data against a bank of theoretically predicted
waveforms to identify astrophysical
signals~\cite{Maggiore:GWbook,Dhurandhar:1994,Allen:findchirp}.
This paradigm carries a fundamental epistemological limitation ---
\emph{we will only see what we expect to see}.
The waveform bank is a projection of our current theoretical understanding
of general relativity and compact binary dynamics;
signals that do not conform to those expectations pass undetected.

This limitation motivates a two-step program.
\emph{Step one}, the subject of this paper, is to replace the
expensive direct evaluation of NR surrogate waveforms with a
fast, differentiable neural surrogate that is faithful to the
current theoretical model yet orders of magnitude cheaper.
This frees the computational budget
for \emph{step two}: learning the distribution of GW signals
from data alone, without anchoring to any specific theoretical template,
and thereby enabling the detection of signals outside our
current theoretical imagination.
Paradoxically, by fully embracing modern machine learning,
we may return to an older mode of science ---
where experiment precedes theory, and anomalies are discovered
before they are explained~\cite{Modrekiladze:GW_AI:2024}.
We reserve step two for a companion paper; here we focus on
building the differentiable surrogate that makes it tractable.

Concretely, gravitational-wave parameter estimation (PE) requires
millions of likelihood evaluations per event, each demanding a waveform
at a new point in an 8-dimensional precessing-spin parameter space. Complementary techniques such as relative binning (heterodyning)~\cite{Zackay:2018qdy}
reduce the per-evaluation likelihood cost by exploiting the smoothness of the waveform ratio across the bank; here we instead target the cost of
\emph{generating} each waveform.
For precessing binary black holes (BBH), \NRSur{}~\cite{Varma:2019csw}
provides state-of-the-art accuracy but at a cost of $\sim\!\SI{12}{ms}$
per waveform on a single CPU core --- a practical bottleneck for
real-time inference, population studies, and Fisher-matrix forecasting.

Fast surrogate models based on reduced-basis and reduced-order
methods~\cite{Field:2014,Blackman:2015,OShaughnessy:2017}, now spanning
aligned-spin~\cite{Varma:2018} and fully
precessing~\cite{Blackman:2017a,Blackman:2017b,Varma:2019csw}
NR waveforms, have demonstrated orders-of-magnitude speedups while
retaining sufficient accuracy for PE.
Neural-network surrogates have extended this paradigm further~\cite{Chua:2018woh,Khan:2020fso},
and a complementary line of
work forgoes waveform generation altogether, training neural networks to
estimate the Bayesian posterior
directly~\cite{George:2018,Gabbard:2019,Green:2020,Dax:2021tsq}.
Neural surrogates for precessing-spin waveforms have recently appeared
for effective-one-body (EOB) models~\cite{Thomas:2022,Whittall:2026};
however, no neural surrogate exists for \NRSur{} specifically~---
the highest-fidelity precessing NR-based model~---
and none of these surrogates is differentiable, which precludes
gradient-based samplers and analytic sensitivity studies.

In this work we construct a fully differentiable surrogate for
\NRSur{} covering its full intrinsic parameter space --- the mass ratio
and the two spin vectors, $(q,\,\chi_{1x},\,\chi_{1y},\,\chi_{1z},
\,\chi_{2x},\,\chi_{2y},\,\chi_{2z})$ --- augmented by the reference
orbital frequency $\omega_0$, for eight input dimensions in total.
Crucially, the surrogate targets the \emph{inertial-frame
spherical-harmonic modes} $h_{\ell m}$ rather than a single polarization
at a fixed orientation: an analytic spin-weighted-spherical-harmonic
projection then yields $h_+,h_\times$ at any orientation, so the
parameter-estimation applications --- which marginalize over inclination,
polarization, and sky position --- are physically complete.
Because that projection is analytic and differentiable, gradients are
available with respect to the extrinsic parameters as well as the intrinsic
ones. The model is lightweight ($5.1\times 10^6$ parameters),
runs on commodity GPUs, and is released publicly
with a simple Python interface.

\section{Method}
\label{sec:method}

\subsection{Mode representation and projection}
\label{sec:modes}

To make the surrogate orientation-complete we target the inertial-frame
spherical-harmonic modes $h_{\ell m}(t)$ of \NRSur{} for all
$2\leq\ell\leq4$ ($21$ modes), obtained from
\texttt{get\_td\_waveform\_modes}~\cite{Usman:2016}. Any observed strain
follows from the analytic spin-weighted spherical-harmonic (SWSH)
projection
\begin{equation}
  h_+ - i\,h_\times = \sum_{\ell,m} h_{\ell m}\,
    {}_{-2}Y_{\ell m}(\iota,\varphi),
  \label{eq:projection}
\end{equation}
which we implement as a differentiable PyTorch layer (closed-form Wigner-$d$
SWSHs, validated against \textsc{lal} to $10^{-15}$). Equation~\eqref{eq:projection}
reproduces the \NRSur{} polarizations at arbitrary $(\iota,\varphi)$ to
machine precision, simultaneously fixing the mode-frame, SWSH, and projection
conventions.

\subsection{Per-mode compression}
\label{sec:svd}

The modes are aligned on a common time grid (peak-aligned on the $(2,2)$
amplitude, with a Planck taper at the band-limited start) and compressed
\emph{per mode}. Each oscillatory mode is represented by its amplitude and
unwrapped phase, $h_{\ell m}=A_{\ell m}e^{i\Phi_{\ell m}}$, while the
nonoscillatory $m=0$ memory modes --- for which the phase is ill defined ---
use a real/imaginary representation. Every channel is compressed with its own
truncated singular-value decomposition (SVD), with the number of basis vectors
$n_{\ell m}$ allocated by an amplitude-weighted criterion: a mode receives
basis vectors until its contribution to the total mismatch falls below a fixed
budget, so the dominant $(2,\pm2)$ modes receive ${\sim}160$ while negligible
high-$\ell$ modes receive as few as $2$. This yields a
$3548$-dimensional coefficient vector
$\mathbf{c}\in\mathbb{R}^{3548}$. Encoding, decoding, and the projection of
Eq.~\eqref{eq:projection} are all differentiable, and the compressed modes
reconstruct $h_+$ at arbitrary orientation with mean match $0.996$, confirming
that compression is not the accuracy bottleneck.

\subsection{Network architecture}
\label{sec:arch}

Given the 8-dimensional parameter vector $\lambda$, we first
embed it into a higher-dimensional feature space via a
random Fourier feature (RFF) map~\cite{Rahimi:2007,Tancik:2020}:
\begin{equation}
  \phi(\lambda) = \sqrt{\tfrac{2}{d}}
    \left[\cos(W\lambda),\,\sin(W\lambda)\right]^T\!,
  \quad W_{ij}\sim\mathcal{N}(0,\sigma^2),
\end{equation}
with $d=128$ features and scale $\sigma=1.0$.
This embedding replaces the raw parameters as the network input
and was found to reduce the validation loss by $\sim\!10\%$
compared to $\sigma=8.0$ in a controlled sweep
over five architecture variants.

The embedded features are passed through a shared residual
network~\cite{He:2016} with $n_\mathrm{blocks}=6$ blocks, each
consisting of two fully-connected layers of width $512$ with layer
normalization~\cite{Ba:2016}, GELU activations~\cite{Hendrycks:2016},
and dropout~\cite{Srivastava:2014} ($p=0.10$).
The shared trunk feeds one linear head per mode, whose outputs are
concatenated into the $3548$-dimensional coefficient vector. We deliberately
omit any global output nonlinearity: the coefficients span several orders of
magnitude across the heterogeneous modes, so a single saturating $\tanh$ would
clip the dominant ones. Instead each channel is standardized independently,
which keeps every mode's dynamic range comparable in the loss.
The full architecture contains $5.1\times10^6$ trainable parameters.

\subsection{Training}
\label{sec:training}

We generate $6\times10^5$ \NRSur{} waveforms using
PyCBC~\cite{Usman:2016}, sampling parameters from a Sobol
sequence~\cite{Sobol:1967} over
$q\in[1,4]$, $|\boldsymbol{\chi}_{1,2}|\leq 0.8$,
with the reference orbital frequency $\omega_0$
drawn from the surrogate's own validity range.
Total mass is fixed at $M=\SI{50}{M_\odot}$; the waveform
scales trivially with $M$ in both amplitude and frequency,
so the surrogate covers arbitrary total masses.
The dataset is split 85/15 into training and validation sets.

Training proceeds in two stages. First, the network is pre-trained to
regress the standardized coefficients with the Huber
loss~\cite{Huber:1964} ($\delta=0.5$), using
AdamW~\cite{Kingma:2014,Loshchilov:2017} with learning rate
$3\times10^{-4}$, weight decay $3\times10^{-3}$, and a cosine annealing
schedule~\cite{Loshchilov:2016}. Second --- because the mode-to-strain
projection of Eq.~\eqref{eq:projection} is differentiable --- we
\emph{fine-tune against the match itself}: for each parameter point we sample
random orientations $(\iota,\varphi)$, project both the predicted and the true
modes to $h_+(f)$, and minimize the PSD-weighted mismatch directly. Since the
match is scale-invariant, a logarithmic norm anchor is added to preserve the
overall strain amplitude (without it the amplitude drifts by orders of
magnitude while the match stays high). This second stage targets the reported
metric and supplies most of the final accuracy. Training is performed on a
single NVIDIA RTX 3090.

\section{Results}
\label{sec:results}

\subsection{Waveform accuracy}
\label{sec:accuracy}

We evaluate the trained model on $150$ held-out parameter points
(freshly sampled, disjoint from training) using the PSD-weighted overlap
(``match'')~\cite{Cutler:1994pb} computed with PyCBC~\cite{Usman:2016}
against the \texttt{aLIGOZeroDetHighPower} noise curve~\cite{aLIGO:2015},
with a lower frequency cutoff of $\SI{25}{Hz}$.
All matches are evaluated at the fiducial total mass
$M=\SI{50}{M_\odot}$ of Sec.~\ref{sec:training}: the waveform itself
scales trivially with $M$, but the PSD weighting does not, so the quoted
figures are specific to this fiducial mass and will shift at total masses
for which the detector band samples a different portion of the
finite-length waveform.
We report two complementary
metrics: the \emph{fixed-orientation} match of the face-on $h_+$ (the
quantity reported by single-polarization surrogates) and the
\emph{orientation-averaged} match over eight random orientations
$(\iota,\varphi)$ per point --- the orientation-complete quantity relevant to
parameter estimation. Table~\ref{tab:matches} summarizes the results.

\begin{table}[h]
\centering
\caption{PSD-weighted match for 150 held-out test waveforms, as
fixed-orientation (face-on $h_+$) and orientation-averaged
(8 random $(\iota,\varphi)$ per point) values.}
\label{tab:matches}
\begin{tabular}{lccccc}
\toprule
 & & \multicolumn{2}{c}{Fixed orientation} & \multicolumn{2}{c}{Orient.-averaged} \\
\cmidrule(lr){3-4}\cmidrule(lr){5-6}
Range & $N$ & Mean & Median & Mean & Median \\
\midrule
$q\in[1,4.0]$    & 150 & 0.975 & 0.988 & 0.940 & 0.975 \\
\midrule
$q\in[1,1.75]$   & 44  & 0.988 & 0.992 & 0.983 & 0.986 \\
$q\in[1.75,2.5]$ & 36  & 0.988 & 0.991 & 0.962 & 0.978 \\
$q\in[2.5,3.25]$ & 37  & 0.977 & 0.987 & 0.927 & 0.927 \\
$q\in[3.25,4.0]$ & 33  & 0.941 & 0.957 & 0.874 & 0.866 \\
\bottomrule
\end{tabular}
\end{table}

\begin{figure}[h]
  \centering
  \includegraphics[width=\columnwidth]{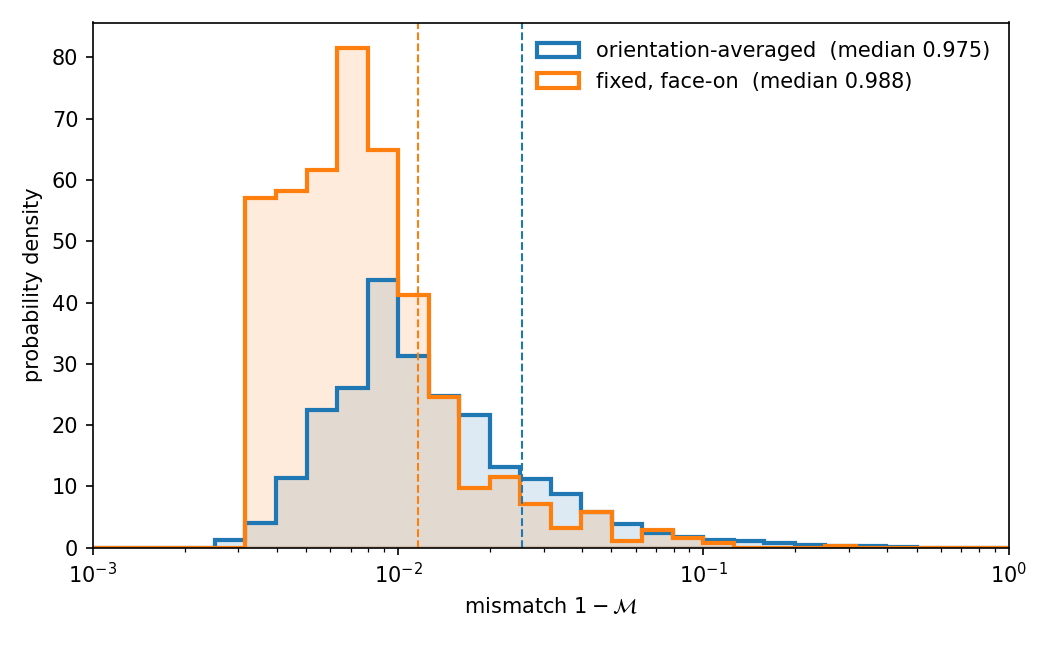}
  \caption{Density-normalized distribution of the PSD-weighted mismatch
           $1-\mathcal{M}$ over the test set: the fixed-orientation match
           (face-on $h_+$, one per point, $N=150$) and the matches pooled over
           the eight sampled orientations per point ($N=8\times150$). The two
           populations are compared by shape rather than raw count; dashed
           lines mark the medians.}
  \label{fig:match_hist}
\end{figure}

Across the full range the fixed-orientation median match ($0.988$) matches
that of an earlier single-polarization version of this surrogate,
while the surrogate now also
reconstructs arbitrary orientations (orientation-averaged median $0.975$). The
strain amplitude is recovered with a median ratio of $0.98$
(10th--90th percentile $[0.81,1.34]$), so the model carries absolute distance
and signal-to-noise information and not only template shape. The match
distribution is shown in Fig.~\ref{fig:match_hist}.

\begin{figure}[t]
  \centering
  \includegraphics[width=\columnwidth]{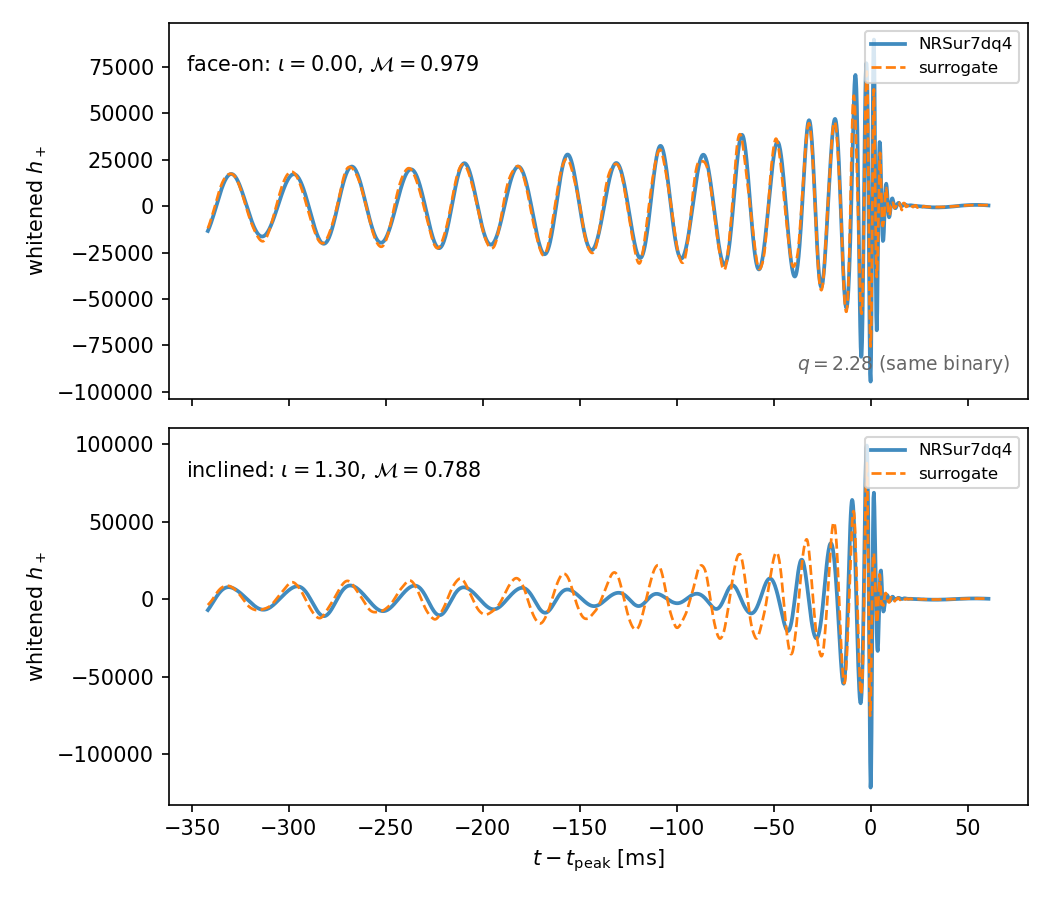}
  \caption{The orientation limitation at the level of the observable strain:
           whitened, peak-aligned $h_+(t)$ (arbitrary units) of the
           \emph{same} binary
           ($q=2.28$) viewed face-on (\textit{top}) and inclined
           ($\iota=1.3$, \textit{bottom}); surrogate (dashed orange) vs.\
           \NRSur{} (solid blue). Face-on, where the strain is carried
           entirely by the $m=2$ modes and dominated by the $(2,2)$
           quadrupole, the surrogate is faithful ($\mathcal{M}=0.979$); viewed
           inclined --- a deliberately unfavorable orientation, drawn from the
           tail of the match distribution at this $q$ --- the less
           well-modeled subdominant modes enter and the agreement degrades
           ($\mathcal{M}=0.788$). Both quoted $\mathcal{M}$ are
           single-orientation matches. This is the strain-level
           origin of the gap between the fixed-orientation and
           orientation-averaged columns of Table~\ref{tab:matches}.}
  \label{fig:orientation}
\end{figure}

Resolving Eq.~\eqref{eq:projection} by mode, the dominant quadrupole
$(2,\pm2)$ and the next-largest $(2,\pm1)$ modes are reproduced with median
mode overlaps $\geq0.94$, while the subdominant $\ell=4$ modes and the $m=0$
memory modes are less well resolved; together the latter carry ${\sim}1\%$ of
the in-band power, so the fixed-orientation strain match remains high. The
consequence is visible directly in the observable strain
(Fig.~\ref{fig:orientation}): viewed face-on the surrogate is faithful, but
viewed at an inclined orientation --- where the subdominant modes contribute
most strongly --- the agreement degrades, which is why the orientation-averaged
match trails the fixed-orientation one. The
degradation at high mass ratio is systematic --- the orientation-averaged mean
falls from $0.983$ at $q<1.75$ to $0.874$ at $q>3.25$ --- consistent with the
precession amplitude growing with both mass ratio and primary
spin~\cite{Zimmerman:2015}; this region, at the boundary of \NRSur{}'s own
validity range, is discussed in Sec.~\ref{sec:discussion}.

\begin{figure}[t]
  \centering
  \includegraphics[width=\columnwidth]{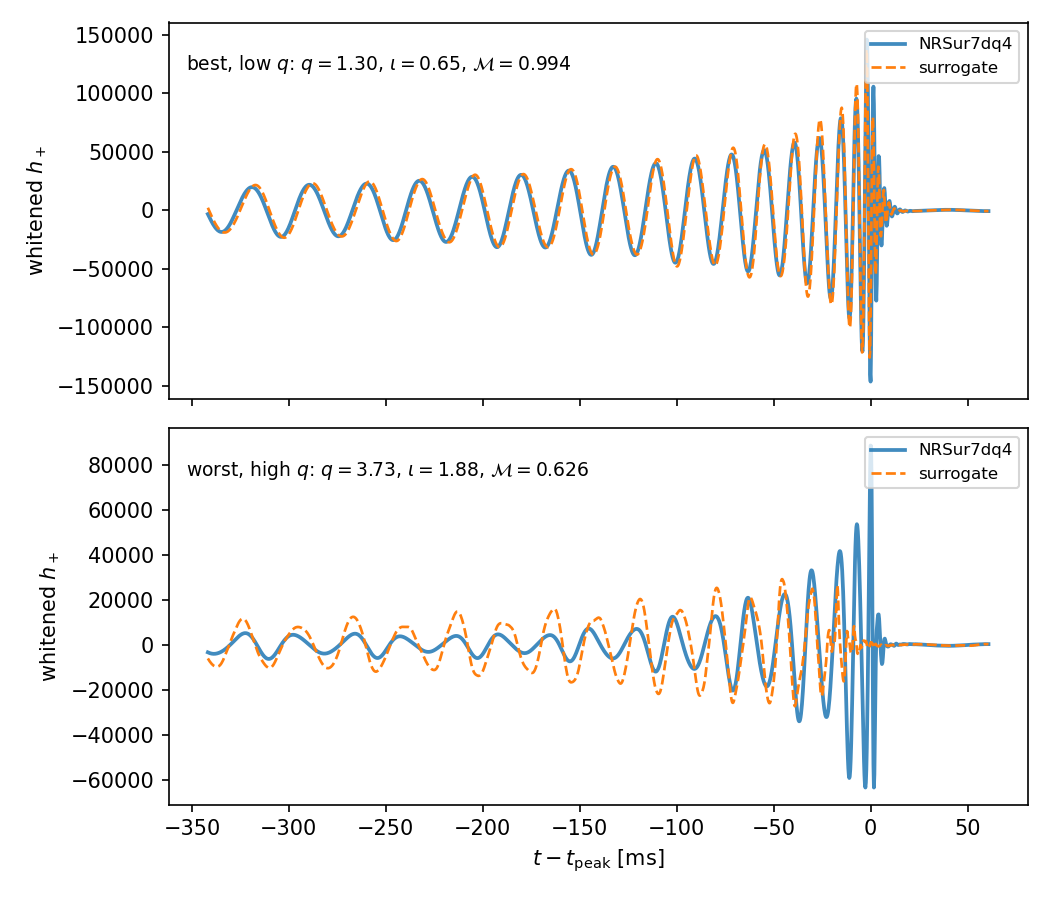}
  \caption{Whitened, peak-aligned time-domain $h_+(t)$ (arbitrary units):
           surrogate (dashed orange)
           vs.\ \NRSur{} (solid blue), for a clean low-$q$ case (\textit{top})
           and a high-$q$ worst case (\textit{bottom}); the quoted
           $\mathcal{M}$ are single-orientation matches at the annotated
           inclinations. Whitening by the
           \texttt{aLIGOZeroDetHighPower} PSD weights the picture by where the
           signal-to-noise actually lies, so the visual agreement tracks the
           PSD-weighted match $\mathcal{M}$. The high-$q$ case shows the phase
           drift toward merger that drives the low match.}
  \label{fig:waveform_overlay}
\end{figure}

\subsection{Generation speed}
\label{sec:speed}

Table~\ref{tab:speed} compares the surrogate throughput
(single NVIDIA RTX 3090) against direct \NRSur{} evaluation
(single CPU thread), at $M=\SI{50}{M_\odot}$.

\begin{table}[h]
\centering
\caption{Waveform generation throughput, including the full mode decode and
SWSH projection to $h_+(f)$.}
\label{tab:speed}
\begin{tabular}{lrr}
\toprule
Method & Latency (ms) & Throughput (wf/s) \\
\midrule
\NRSur{} (CPU)        & $\sim\!12$   & $\sim\!82$ \\
\midrule
Surrogate, batch 1    & 12.0 & $8.4\times10^1$ \\
Surrogate, batch $10^2$ & 12.6 & $7.9\times10^3$ \\
Surrogate, batch $10^3$ & 28   & $3.5\times10^4$ \\
\bottomrule
\end{tabular}
\end{table}

In the batched regime relevant to nested sampling and population inference ---
where the likelihood demands many waveforms in parallel --- the surrogate
reaches ${\sim}3.5\times10^4$ waveforms per second, about $430\times$ faster
than direct \NRSur{} evaluation. This is the operating point that matters for
inference, and it is achieved while delivering both polarizations at arbitrary
orientation and exact gradients, neither of which the direct model provides.
Each waveform now entails decoding $21$ time-domain modes and the SWSH
projection of Eq.~\eqref{eq:projection}, so the model is heavier than a
single-polarization surrogate: the single-waveform latency ($\SI{12}{ms}$) is
comparable to direct evaluation and is dominated by fixed GPU kernel-launch
overhead --- batches of $1$ and $10^2$ take essentially the same wall time ---
so the per-waveform cost falls steeply with batch size. The reconstructed
time-domain modes form the dominant memory cost, so the largest batches are
tiled rather than evaluated monolithically.

\subsection{Differentiability}
\label{sec:diff}

Because every operation in the surrogate pipeline
(Fourier embedding, ResNet layers, SVD reconstruction)
is differentiable, the gradient
$\partial\tilde{h}/\partial\lambda$
is available via automatic differentiation~\cite{Baydin:2018}
at negligible additional cost.
This enables two applications not possible with
tabulated surrogates such as \NRSur{}:

\textit{Fisher-matrix estimation.}
The Fisher information matrix~\cite{Cutler:1994pb,Poisson:1995ef},
which underlies fast forecasting tools such as
gwbench~\cite{Borhanian:2020},
\begin{equation}
  \Gamma_{ij} = \left\langle
    \frac{\partial\tilde{h}}{\partial\lambda_i}
    \Bigg|
    \frac{\partial\tilde{h}}{\partial\lambda_j}
  \right\rangle
\end{equation}
requires the gradient of the strain with respect to every parameter.
Because the mode-to-strain projection of Eq.~\eqref{eq:projection} is analytic,
the surrogate is differentiable not only in the eight intrinsic parameters ---
whose derivatives propagate through the network and the per-mode decode --- but
also in the extrinsic parameters $(\iota,\varphi,\psi,\ln d_L,t_c)$, whose
derivatives flow through the projection and the standard distance and
time-shift factors. This yields a full $13$-dimensional Fisher matrix from
automatic differentiation~\cite{Paszke:2019}, a capability that tabulated
surrogates such as \NRSur{} cannot provide for the extrinsic parameters
analytically. We validated the autodiff derivatives against finite differences,
with per-parameter agreement above $0.9999$; the primary advantage of AD is
\emph{exactness}: no step-size selection, no truncation error, and no numerical
cancellation near singular points of the waveform. The resulting Fisher matrix
inverts cleanly, its conditioning set by the expected hierarchy between
well-constrained parameters (chirp mass, time of coalescence) and the
weakly-constrained in-plane spins.

\textit{Gradient-based PE.}
The more significant benefit of differentiability is for
gradient-based sampling.
Hamiltonian Monte Carlo (HMC)~\cite{Duane:1987,Neal:2011} and
NUTS~\cite{Hoffman:2014} require
$\nabla_\lambda \log \mathcal{L}$ at every leapfrog step,
obtained here via a single reverse-mode backward pass ($\sim\!\SI{20}{ms}$),
and now over the full intrinsic-plus-extrinsic parameter set.
Two properties make exact gradients preferable to finite differences in
this setting.
First, \emph{exactness}: noisy finite-difference gradients corrupt the
symplectic integrator, degrading the proposal acceptance rate and
effectively lengthening the chain --- the benefit is not merely tidiness
but the correctness of the sampler.
Second, \emph{scaling}: reverse-mode cost is $\mathcal{O}(1)$ in
$n_\lambda$, whereas finite differences scale as $\mathcal{O}(n_\lambda)$.
At $n_\lambda = 8$ the wall-clock difference is small, but it becomes
material over the full $13$-dimensional intrinsic-plus-extrinsic space the
surrogate already exposes, and when it is embedded in a higher-dimensional
inference framework.
This makes gradient-based PE directly compatible with
samplers such as \texttt{numpyro}~\cite{Phan:2019} and
\texttt{flowMC}~\cite{Wong:2022}
without modification, potentially reducing the number of
likelihood evaluations by orders of magnitude compared to
random-walk MCMC.
We demonstrate this end-to-end in Appendix~\ref{app:pe}: a gradient-based
NUTS recovery on a zero-noise injection of the \emph{true} \NRSur{} waveform,
sampled with \texttt{pyro}~\cite{Bingham:2019}, recovers the intrinsic
parameters $(q,\chi_{1z},\chi_{2z})$ without significant bias
($\leq\!0.3\sigma$ at SNR $25$) and exposes only a mild inclination--distance
pull that is consistent with the reduced fidelity of the subdominant modes
(Sec.~\ref{sec:discussion}). This both establishes that the surrogate's
gradients drive a sampler end-to-end and provides a first check that its
waveform error does not bias parameter recovery. A full injection campaign
over noise realizations and the complete precessing parameter space ---
the inference study proper --- is reserved for the companion
work~\cite{Modrekiladze:GW_AI:2024}.

\section{Discussion}
\label{sec:discussion}

\paragraph{High-$q$ regime.}
A limitation of the current model is its accuracy at high mass ratio: the
orientation-averaged mean match falls from $0.983$ at $q<1.75$ to $0.874$ at
$q>3.25$. This region corresponds to the boundary of \NRSur{}'s own validity
range ($q\leq4$) and likely reflects a combination of (i)~reduced training-data
density near $q=4$ under Sobol sampling, (ii)~increased waveform complexity
(strong precession) at high $q$, and (iii)~potential accuracy degradation of
\NRSur{} itself near its validity boundary. Augmenting the training set with a
focused sample of $q\in[3.5,4]$ waveforms is expected to help.

\paragraph{Subdominant modes.}
The dominant accuracy limitation specific to the mode model is the fidelity of
the subdominant modes: while the $(2,\pm2)$ and $(2,\pm1)$ modes reach median
overlaps $\geq0.94$, the $\ell=4$ and $m=0$ memory modes are learned only at the
$0.1$--$0.3$ level. They carry ${\sim}1\%$ of the in-band power, so the
fixed-orientation match is largely unaffected, but they contribute more strongly
at inclined orientations and are the main reason the orientation-averaged match
trails the fixed-orientation one. We observed steady improvement with dataset
size (the $(2,\pm1)$ overlap rose from $0.24$ to $0.94$ between $2.5\times10^5$
and $6\times10^5$ training waveforms), suggesting the subdominant modes are
data-limited rather than capacity-limited. A coprecessing-frame representation,
which factors out the orbital precession before modeling the modes, is the
natural next step and is expected to sharpen these modes substantially.

\paragraph{Comparison to existing surrogates.}
Neural-network surrogates for precessing-spin EOB models have been
presented by Thomas et al.~\cite{Thomas:2022} (SEOBNRv4PHM) and
Whittall \& Pratten~\cite{Whittall:2026} (SEOBNRv5PHM, $q\leq10$,
with demonstrated Bayesian PE on real events).
To our knowledge, no neural surrogate has been published for
\NRSur{} specifically~--- an NR-based model whose accuracy ceiling
lies above that of any current EOB approximant in the precessing
regime~\cite{Varma:2019csw}~--- nor are any of these surrogates
differentiable with respect to the physical parameters.
The aligned-spin neural surrogates of~\cite{Chua:2018woh,Khan:2020fso}
also do not cover the full precessing-spin space.
A distinct recent direction uses agentic large language models to
construct GW models: the GWAgent framework~\cite{GWAgent:2605.11280}
builds interpretable \emph{analytic} surrogates --- demonstrated on
\textsc{SEOBNRv5EHM}~\cite{Gamboa:2024}, a different (eccentric,
aligned-spin) approximant --- while the gwBenchmarks
suite~\cite{gwBenchmarks:2605.11269} stress-tests such agents on GW
modeling tasks.
GWAgent reports a median mismatch of $6\times10^{-4}$, below our
fixed-orientation median mismatch of $\sim\!1\times10^{-2}$; its
$\sim\!\SI{13}{ms}$ per-waveform cost is comparable to ours for a single
waveform, though our model amortizes to $\sim\!3.5\times10^4$ per second in
batches. These figures are not directly comparable: the two works target
different approximants and parameter spaces, GWAgent optimizes for analytic
interpretability, and our surrogate uniquely provides orientation-complete,
both-polarization, differentiable waveforms.

\paragraph{Outlook.}
The surrogate presented here is a first step toward
fully differentiable, GPU-native GW inference pipelines.
In a companion paper, we train a generative model directly on the
distribution of GW signals --- without anchoring to any fixed
theoretical template --- and ask whether such a model can produce
waveforms that fall \emph{outside} the support of current
NR surrogates~\cite{Modrekiladze:GW_AI:2024,Modrekiladze:transfer:2024}.
In the worst case, the model rediscovers general relativity,
reproducing the known waveform manifold from data alone.
In the best case, it surfaces morphologies that point to
new physics beyond our current theoretical expectations.
The fast, differentiable surrogate developed here provides both
the training signal and the computational infrastructure for
that investigation.

\section{Conclusion}
\label{sec:conclusion}

We have presented a $5.1\times10^6$-parameter neural-network
surrogate for \NRSur{} precessing BBH waveforms covering
its full intrinsic parameter space
$(q,\,\boldsymbol{\chi}_1,\,\boldsymbol{\chi}_2)$
together with the reference frequency $\omega_0$ ---
eight inputs in total.
By predicting the inertial-frame spherical-harmonic modes rather than a
single polarization, the model reconstructs both polarizations at any
orientation through an analytic, differentiable projection. It achieves a
fixed-orientation median match of $0.988$ and an orientation-averaged median
match of $0.975$ over $q\in[1,4]$ while keeping the overall strain amplitude
physical (median ratio $0.98$), generates waveforms at ${\sim}3.5\times10^4$~wf/s in batches
(about $430\times$ faster than direct evaluation), and is differentiable with
respect to both intrinsic and extrinsic parameters --- enabling a full
$13$-parameter Fisher matrix and gradient-based sampling.
These properties make it suitable as a drop-in replacement
for \NRSur{} in nested-sampling
PE~\cite{Skilling:2006,Veitch:2015,Ashton:2019,Speagle:2020},
Fisher-matrix studies, and gradient-based inference pipelines.
Code and trained weights are available at
\href{https://github.com/solipsism1/AIGWsur}{\texttt{github.com/solipsism1/AIGWsur}}.

\medskip
\noindent\emph{Note added.}---While finalizing this manuscript we became aware of
the concurrent work of P\"urrer \emph{et al.}~\cite{Puerrer:2026}, a
neural-network surrogate of \NRSur{} with a differentiable \textsc{JAX}
waveform-to-likelihood pipeline, which overlaps with some of our results. The two
constructions are architecturally complementary: Ref.~\cite{Puerrer:2026} mirrors
\NRSur{}'s internal coprecessing decomposition with a bank of 25 networks
($4.8\times10^{7}$ parameters), whereas we model the inertial-frame radiation
field directly with a single compact network ($5.1\times10^{6}$ parameters).
Beyond this, the present work differs in two respects. First, our surrogate is
trained \emph{through} its own differentiable mode-to-strain projection,
fine-tuning directly on the PSD-weighted match (Sec.~\ref{sec:training}) rather
than on a node-level regression loss. Second, we demonstrate gradient-based
parameter estimation end to end: the NUTS recovery of Appendix~\ref{app:pe} draws
$\nabla_{\lambda}\log\mathcal{L}$ from reverse-mode automatic differentiation at
every leapfrog step, whereas the inference demonstrations of
Ref.~\cite{Puerrer:2026} employ gradient-free nested sampling. Conversely, their
coprecessing-frame representation attains higher waveform fidelity, supporting
the expectation of Sec.~\ref{sec:discussion} that factoring out precession before
learning is the natural route to sharpening the subdominant modes; we view the
two works as convergent evidence that fast, differentiable surrogates of \NRSur{}
are the right infrastructure for gradient-based gravitational-wave inference.

\begin{acknowledgments}
We thank Rafael Porto and Matias Zaldarriaga for useful comments and discussions.
\end{acknowledgments}

\appendix

\section{Gradient-based recovery on a synthetic injection}
\label{app:pe}

To verify that the surrogate's exact gradients drive a sampler end to end ---
and, more importantly, that its waveform error does not bias parameter
recovery --- we perform a controlled injection study. We inject the
\emph{true} \NRSur{} waveform (not a surrogate draw) at
$q=2$, $\boldsymbol{\chi}_1=(0.10,0.05,0.30)$,
$\boldsymbol{\chi}_2=(-0.10,0.05,-0.20)$, inclination $\iota=0.9$,
and recover it with the surrogate likelihood. The injection is performed in
zero noise (an ``Asimov'' injection), so that any offset of the posterior
from the injected values is a \emph{pure waveform systematic} rather than a
noise fluctuation. Crucially, the injected data and the recovery template are
projected to detector strain through the \emph{same} analytic SWSH layer of
Eq.~\eqref{eq:projection}; only the mode content differs --- true \NRSur{}
modes for the data, network-predicted modes for the template --- so the
experiment isolates the network's intrinsic mode error as the sole possible
source of bias.

The single-detector strain is whitened by the
\texttt{aLIGOZeroDetHighPower} PSD and scaled to a single-detector
signal-to-noise ratio (SNR) of $25$. We sample the Gaussian likelihood with the
No-U-Turn sampler (NUTS)~\cite{Hoffman:2014} as implemented in
\texttt{pyro}~\cite{Bingham:2019}, drawing the gradient
$\nabla_\lambda\log\mathcal{L}$ at each leapfrog step from a single
reverse-mode pass through the surrogate. We sample the intrinsic
parameters that dominate the phasing,
$(q,\chi_{1z},\chi_{2z})$, together with the inclination, luminosity
distance, coalescence time, and the azimuthal phase of
Eq.~\eqref{eq:projection},
$(\iota,\ln d_L,t_c,\varphi)$; the in-plane spins and polarization are held
at their injected values for this demonstration. The sampler reaches a mean
acceptance probability of $0.9$ against the target value of $0.8$, with stable
step-size adaptation.

\begin{figure}[t]
  \centering
  \includegraphics[width=\columnwidth]{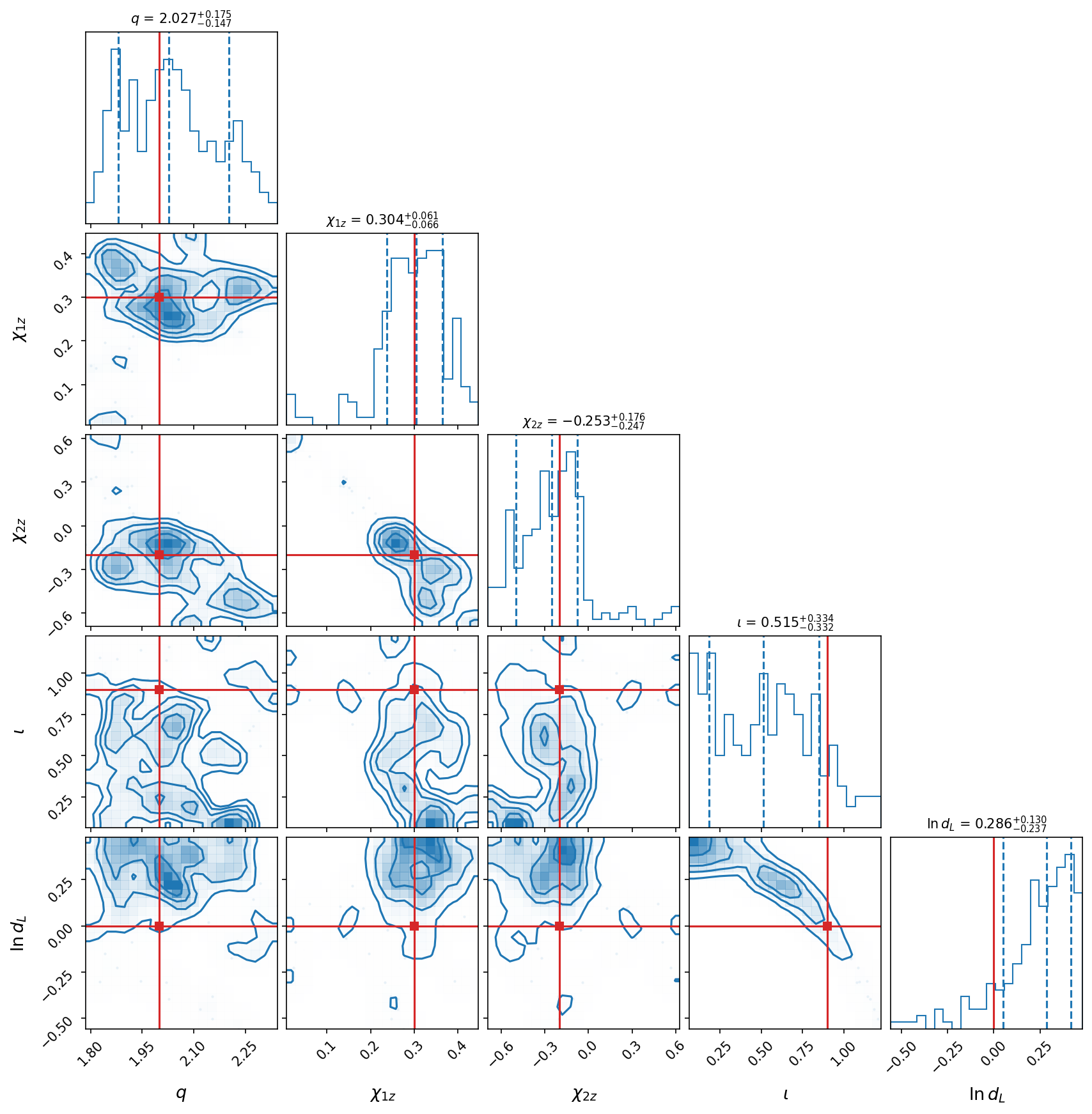}
  \caption{Marginal posteriors from the gradient-based (NUTS) recovery of a
           zero-noise injection of the true \NRSur{} waveform, recovered with
           the surrogate likelihood at single-detector SNR $25$. Red lines
           mark the
           injected values. The intrinsic parameters $(q,\chi_{1z},\chi_{2z})$
           are recovered without significant bias ($\leq\!0.3\sigma$ at this
           SNR); the
           inclination and distance show a mild ($\sim\!1\sigma$),
           anti-correlated pull along the standard distance--inclination
           degeneracy. The coalescence time and reference phase (nuisance
           parameters) are marginalized and not shown.}
  \label{fig:pe_corner}
\end{figure}

The recovered posteriors are shown in Fig.~\ref{fig:pe_corner}. The intrinsic
parameters are recovered without significant bias: the posterior means lie
within $0.26\sigma$ ($q$), $0.05\sigma$ ($\chi_{1z}$), and $0.15\sigma$
($\chi_{2z}$) of the injected values. The inclination and distance are pulled
by $\sim\!1\sigma$ ($-1.3\sigma$ and $+1.1\sigma$ respectively) in an
anti-correlated fashion along the well-known distance--inclination
degeneracy. This is the expected place for a residual systematic to appear:
the inclination is constrained largely by the relative amplitudes of the
subdominant modes against the dominant $(2,2)$, and those modes
(the $\ell=4$ and $m=0$ memory modes) are exactly the channels the surrogate
reproduces least accurately (Sec.~\ref{sec:discussion}). The phase $\varphi$
absorbs a constant phase-convention offset between the surrogate and \NRSur{}
and is treated as a nuisance.

Because the injection is noise-free, these offsets are pure waveform
systematics, fixed in absolute parameter units: $\Delta q=+0.035$,
$\Delta\iota=-0.37$~rad, and $\Delta\ln d_L=+0.23$, with
$\Delta\chi_{1z}=-0.004$ and $\Delta\chi_{2z}=-0.036$.
The statistical width, by contrast, contracts as $1/\mathrm{SNR}$, so the
$\sigma$-normalized figures above are specific to SNR $25$: the same waveform
error that produces a ${\sim}1\sigma$ inclination--distance pull here
corresponds to ${\sim}3\sigma$ at SNR $75$. The growth of this systematic
with loudness is a further, quantitative motivation for the
subdominant-mode improvements of Sec.~\ref{sec:discussion}.

We emphasize that this is a single zero-noise injection in a reduced parameter
set, intended to establish that gradient-based inference runs end to end on
the surrogate and that the dominant intrinsic parameters are recovered without
bias. A full injection campaign --- many parameter points, noise realizations,
the complete $13$-dimensional precessing-plus-extrinsic space, and a
quantitative comparison of sampler efficiency against nested sampling --- is
the subject of the companion work~\cite{Modrekiladze:GW_AI:2024}.

\newpage
\bibliography{refs}

\end{document}